\documentclass{mem}
\usepackage{natbib}
\usepackage{txfonts}
\usepackage{balance}
\usepackage{graphicx}
\usepackage[a4paper,breaklinks]{hyperref}
\idline{75}{282}

\begin{document}
\def\teff{$T\rm_{eff }$}
\def\kms{$\mathrm {km s}^{-1}$}

\title{
Polar rings dynamics in the triaxial dark matter halo
}

\subtitle{}

\author{
S. \,Khoperskov\inst{1}, A. \, Moiseev\inst{2}
\and A. \, Khoperskov\inst{3}
          }

\offprints{S. Khoperskov}

\institute{
Institute of Astronomy, Russian Academy of Sciences, Pyatnitskaya st., 48, 119017, Moscow, Russia \email{khoperskov@inasan.ru}
\and
Special Astrophysical Observatory, Russian Academy of Sciences, Nizhnii Arkhyz, Karachai-Cherkessian Republic, 357147, Russia
\and
Volgograd State University, Universitetsky pr., 100, 400064, Volgograd, Russia
}

\authorrunning{S. Khoperskov}

\titlerunning{Polar rings dynamics in triaxial DM halo }

\abstract{
Spectroscopic observations at the Russian 6-m telescope  are used to study the two polar ring galaxies (PRGs) from the catalogue by Moiseev et al.: SPRC-7 and SPRC-260. We have analyzed the kinematics of the stellar component of the central galaxies as well as the ionized gas kinematics in the external ring structures. The disc-halo decomposition of rotation curves in two perpendicular directions are considered. The observed 2D velocity fields are compared with the model predictions for different dark halo shapes. Based on these data, we constrain that for potential of DM halo semiaxis ratios  is $s=0.8$, $q=1$ for SPRC-7 and  $s=0.95$, $q=1.1$ for SPRC-260. Using 3D hydrodynamic simulations we also study the dynamics and evolution of the polar component in the potential of the galactic disc and dark halo for these two galaxies. We show that the polar component is dynamically quasi-stable on the scale of $\sim10$ dynamical times (about a few Gyr). This is demonstrate the possibility for the growth of a spiral structure, which then steadily transforms to a lopsided gaseous system  in the polar pane.
}
\maketitle{}

\section{Introduction}
The modern $\Lambda CDM$ cosmological theory predicts the presence of  dark matter in the vicinity of galaxies. Which leads to the fact that the rotation of disc galaxies is partly supported by the massive dark matter
component~\citep{Athanasoula-Bosma-Papaioannou-1987!Halo-spiral-gal-swing, Bottema1993}. Cosmological simulations also predict a non-spherical
density distribution in DM haloes at $z=0$~\citep{Allgood-etal-2006!shape-halo-N-body, Hayashi-Navarro-2006!kinematics-disc-triaxial-halo}. On the other hand,  the observational constraints on the galactic halo shapes are coming from the gaseous and stellar kinematics in the disc galaxies~\citep{Merrifield2002}. Polar ring galaxies (PRGs) are unique objects for the investigation of  DM halo shapes because the rotation of the matter happens in  two perpendicular directions~\citep{Sackett-Pogge-1995!Polar-ring,Combes-2006!Polar-Ring}. In this letter we have obtained the constrains on the shape of dark matter for two PRGs using the observations of the rotation curves of host S0 galaxies and polar gaseous components.

\begin{figure*}[!t]
\resizebox{\hsize}{!}{\includegraphics[clip=true]{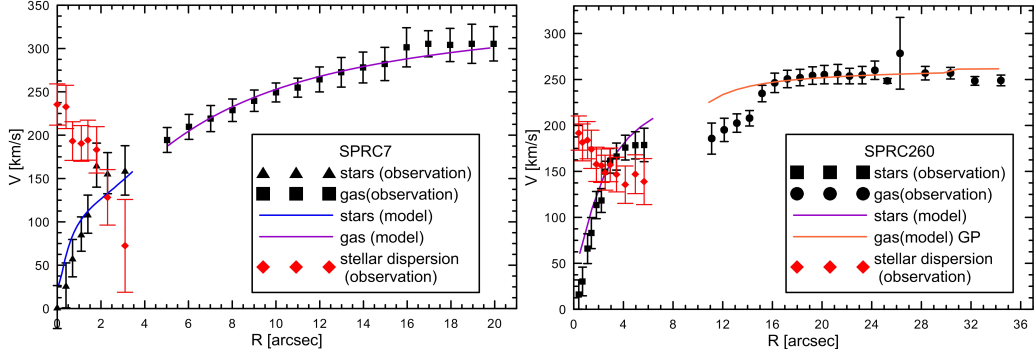}}
\caption{\footnotesize Observation data of the rotation in the galaxy  and in the polar ring planes for both galaxies. Best fit model rotation curves are shown by solid lines. The observed line-of-sight velocity dipsersion distributions of stars (red symbols) were also used in the  simulations. }
\label{fig::compare07}
\end{figure*}

\section{Observations}
Spectroscopic observations at  the 6-m telescope of the SAO RAS were used to study two polar ring galaxies from the catalogue by \citet{moisav2011}: SPRC-7 and SPRC-260. SPRC-7 is an inclined system with the relative
angle $\vartriangle i = 70^\circ$ towards  the central galaxy, meanwhile SPRC-260 is a classic polar case
with $\vartriangle i = 89^\circ$. The observations were carried out with the multi-mode focal reducers SCORPIO \citep{AM05} and SCORPIO-2 \citep{AM12} in the scanning Fabry-Perot
interferometer and long-slit spectroscopic modes. The results of observations are
\begin{enumerate}
\item The line-of-sight velocity and velocity dispersion distributions of the stellar
component along the central galaxy major axis;
\item  The ionized gas velocity fields for the polar component. The data for
SPRC-7  were already presented in \citet{Brosch2010}.
\end{enumerate}

\section{Simulations}
We have simulated the dynamics of the polar gaseous ring solving 3D
hydrodynamical equations in the cylindrical coordinates $(r,\varphi,z)$ with the TVD MUSCL scheme, but neglecting the selfgravitation of gas. The ring
was orientated in the plane $z=0$. The external potential consists of a triaxial isotermal DM halo~\citep{Khoperskov2012} and a contribution from the flattened S0 galaxy. It is well known that the disc perturbs the velocity field of the polar ring~\citep{Theis2006}. Thus it is very important to  carefully estimate the vertical structure of  the galaxy potential for the dynamics of the polar component.  We calculate the potential of the galaxy by TreeCode method~\citep{BarnesHut1986} using an adequate distribution of test particles, reproducing the structure of the central S0 galaxy. This technique allows us to estimate the correct potential far away from the disc plane.
\begin{figure}[!h]
\resizebox{\hsize}{!}{\includegraphics[clip=true]{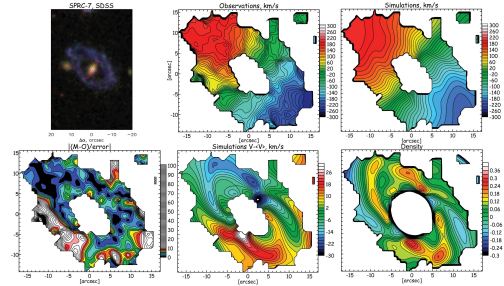}}
\caption{\footnotesize SPRC-7. The top row:   an optical  SDSS image (left);   the ionized gas line-of-sight velocity field $V_{obs}(x,y)$ taken from observations (middle);  the best-fit model $V_{mod}(x,y)$ of the  velocity field  (right).
The bottom row: the difference between $V_{obs}$ and $V_{mod}$ normalized to  the observational errors (left);  the velocity perturbation  $V_{mod} - \langle V_{mod}\rangle$ (middle); the simulated column density of  gas in the polar component at $5$ rotational periods (right). }
\label{fig::compare07}
\end{figure}

Initially, the gas dynamic simulations started from the observed kinematics of polar
components (fig.~1). For both galaxies the simulated velocity field generally agrees  with the observations
 (see Figs.~2 and 3) during many rotation periods (period is $T \sim 300 \div 400$ Myr). This systematic deviation could
be associated with the polar ring warping (fig.~2). Relative velocity perturbations correlate with
the features in the spacial density distribution. However, the polar component is unstable
towards the formation of some large-scale clumps. The maxima of density distributions are
situated in the  regions of   intersection of the polar component plane with the host galaxy rotation
plane. These points are also prominent   on the maps of velocity perturbation $V-\langle V \rangle$
(where $\langle V \rangle$ is the azimuthal averaged velocity).
The results of simulations of SPRC-260 do not allow  to determine any morphological
features on the velocity maps. Despite this fact, the simulated and observed maps are quite
similar. We showed that the polar component is dynamically stable on the scales of about $10$ dynamical times (about a few Gyr).

\begin{figure}[!h]
\resizebox{\hsize}{!}{\includegraphics[clip=true]{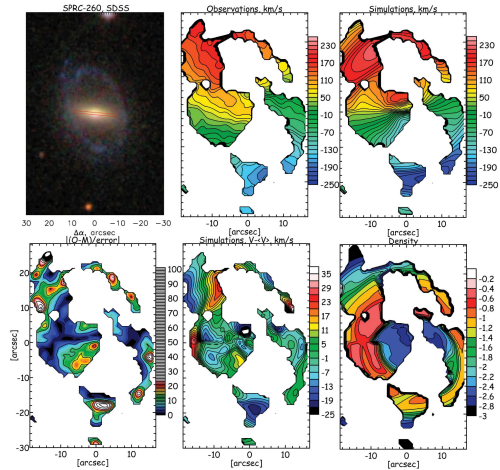}}
\caption{\footnotesize Figures are the same as in figure~\ref{fig::compare07} for SPRC-260, respectively.}
\label{fig::compare260}
\end{figure}
The observed 2D velocity fields for both PRGs were compared with the model predictions for
different dark halo shapes at different times. We have build a $\chi^2_{ring}$-distribution for the observed and simulated velocity fields of the polar gaseous component, depending on time and the halo shape parameter $s=c/b$; the other $\chi^2_{gal}$-distribution for the host galaxy stellar kinematics is  also dependent on time and $q=a/b$ (where $a,b,c$ are the semiaxises of the DM halo potential). We have found that there are no significant time-dependent variations of $\chi^2$ up to $10$ rotation periods and it is increasing  slightly.

\begin{figure*}[t!]
\resizebox{\hsize}{!}{\includegraphics[clip=true]{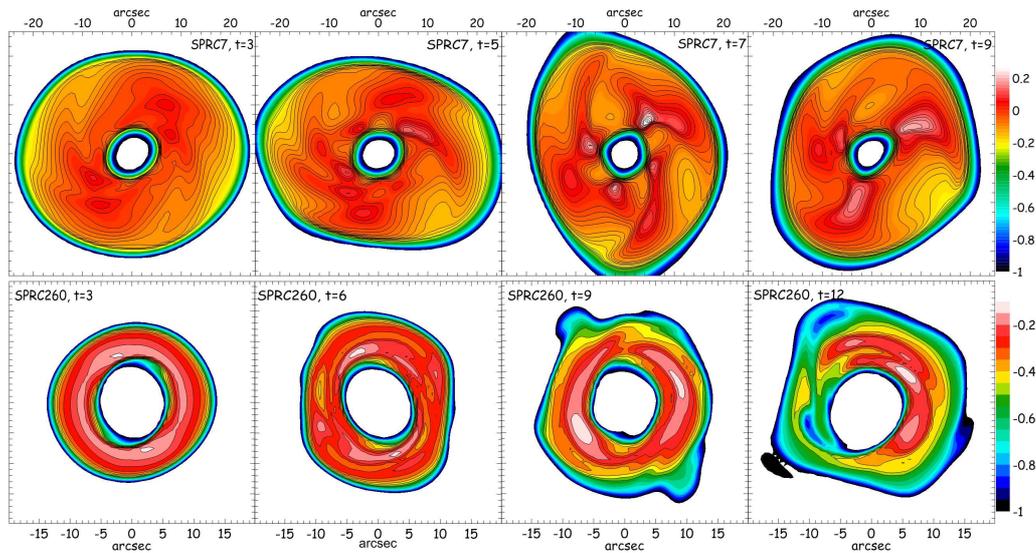}}
\caption{\footnotesize The time-depend evolution of the column density of polar components for SPRC-7 (upper panel) and SPRC-260 (down panel). Time unit corresponds to the one rotation period of the outer side of thering. }
\label{fig::evolution}
\end{figure*}

Detailed simulations are able to investigate the internal evolution of gas in the ring
components. An interaction with a non-axisymmetrical perturbation from the host
galaxy and the triaxial DM halo leads to the formation of spiral structures in the rings. There
is a 2-arm tightly wound spiral structure formation in SPRC-260, which
transforms to a lopsided system. For SPRC-7, multiple arms formed with a large pitch
angle (fig.~4).

The best fits (minima of $\chi^2_{ring}$ distributions) pointed out the oblate character of the DM halo potential distribution in the polar plane for both galaxies: $s = 0.8$ for SPRC-7 and $s=0.95$ for SPRC-260. For the halo  density distribution $q$ and $s$ should be significantly smaller.

\section{Results and discussions}

We propose an oblate DM halo potential for SPRC-7 with the
semiaxis relations: $s=0.8$, $q=1$ and a triaxial halo for SPRC-260: $s=0.95$, $q=1.1$. Note  that this halo shape contradicts with a previous conclusion by \citet{iod} based on the investigation of the polar ring galaxies properties from the  Tully-Fisher relation: ``a flattened polar halo, aligned with the polar ring''. But an oblate halo is strongly preferred over a prolate one from the estimations of dSgr dynamics
$s=0.90-0.95$~\citep{johnston}. Numerous cosmological simulations also predict an oblate DM halo shape  \citep[e.g.][]{kuhlen07}.

The evolution of  morphology for both polar components is quite  different, but there is a
common feature, namely, the   formation of spirals,   which could be determined  from the velocity and
density distributions. A spiral structure forms in the non-self-gravitating gas due to the interactions with the  non-axisisymmetric perturbations of the gravitational potential (DM halo and SO galaxy).  However, the formations on nonlinear large-scale spirals in the polar component require to take into account the self-gravity of the gas and stellar fractions of the polar component.

\begin{acknowledgements}
This work was partly supported by the grants RFBR (12-02-31452, 13-02-00416) and by the RAS program of the  fundamental investigation OFN-17 ``Active processes in galactic and extragalactic objects''.  S.~Khoperskov expresses his gratitude to nonprofit  ``Dynasty'' foundation for financial support. A.~Moiseev also thank the Springer \& EAS granted his participation on EWASS-2012. Numerical simulations were run on supercomputers of NIVC MSU ``Lomonosov'' and ``Chebyshev''.
\end{acknowledgements}

\bibliographystyle{aa}

\begin{thebibliography}{}


\bibitem[\protect\citeauthoryear{Afanasiev \& Moiseev}{2005}]{AM05} Afanasiev V.L., Moiseev A.V.: 2005, AstL 31, 194

\bibitem[\protect\citeauthoryear{Afanasiev \& Moiseev}{2012}]{AM12} Afanasiev V.L., Moiseev A.V.: 2012, Baltic Astronomy 20, 363

\bibitem[\protect\citeauthoryear{Allgood et al.}{2006}]{Allgood-etal-2006!shape-halo-N-body}
   Allgood B., Flores R.A., Primack J.R.,  et al.: 2006, MNRAS, 367, 1781


\bibitem[\protect\citeauthoryear{Athanasoula et al.}{1987}]{Athanasoula-Bosma-Papaioannou-1987!Halo-spiral-gal-swing} Athanasoula E., Bosma A., Papaioannou S.: 1987, A\&A 179, 23

 \bibitem[\protect\citeauthoryear{Barnes and Hut}{1986}]{BarnesHut1986} Barnes J., Hut P.: 1986, Nature, 324, 446

 \bibitem[\protect\citeauthoryear{Bottema}{1993}]{Bottema1993} Bottema, R.: 1993, A\&A 275, 16

\bibitem[\protect\citeauthoryear{Brosch et al.}{2010}]{Brosch2010} Brosch N., Kniazev A.Y., Moiseev A., Pustilnik S.A.: 2010, MNRAS 401, 2067

 \bibitem[\protect\citeauthoryear{Combes}{2006}]{Combes-2006!Polar-Ring}
Combes F., 2006, EAS Publ. Ser., 20, 97

\bibitem[\protect\citeauthoryear{Hayashi and Navarro}{2006}]{Hayashi-Navarro-2006!kinematics-disc-triaxial-halo}
 Hayashi E., Navarro J.F., 2006, MNRAS, 373, 1117

\bibitem[\protect\citeauthoryear{Iodice et al.}{2003}]{iod}
Iodice E., Arnaboldi M., Bournaud F., et al.: 2003, ApJ 585, 730

\bibitem[\protect\citeauthoryear{Johnston et al.}{2005}]{johnston} Johnston K.V., Law D.R., Majewski S.R.: 2005, ApJ 619, 800

\bibitem[\protect\citeauthoryear{Khoperskov et al.}{2012}]{Khoperskov2012} Khoperskov A.V., Eremin M.A., Khoperskov S.A., Butenko M.A., Morozov A.G.: 2012, ARep 56, 16

\bibitem[\protect\citeauthoryear{Kuhlen et al.}{2007}]{kuhlen07} Kuhlen M., Diemand J., Madau P.: 2007, ApJ  671, 1135
\bibitem[\protect\citeauthoryear{Merrifield}{2002}]{Merrifield2002} Merrifield, M. R. 2002, in The Shapes of Galaxies and Their Dark Halos, ed. P.
Natarajan (Singapore: World Scientific), 170

\bibitem[\protect\citeauthoryear{Moiseev et al.}{2011}]{moisav2011}
Moiseev A.V., Smirnova K.I., Smirnova A.A., Reshetnikov V.P.: 2011, MNRAS 418, 244

 \bibitem[\protect\citeauthoryear{Sackett and Pogge}{1995}]{Sackett-Pogge-1995!Polar-ring}
 Sackett P.D., Pogge R.W., 1995, AIP Conf. Proc., 336, 141

\bibitem[\protect\citeauthoryear{Theis et al.}{2006}]{Theis2006} Theis Ch., Sparke L., Gallagher J.: 2006, A\&A 446, 905


\end{thebibliography}

\end{document}